\documentclass[english,english,journal,draftcls,onecolumn,12pt]{IEEEtran}
\usepackage[T1]{fontenc}
\usepackage{babel}
\usepackage{amsmath}
\usepackage{graphicx}
\usepackage[unicode=true,
 bookmarks=false,
 breaklinks=false,pdfborder={0 0 0},pdfborderstyle={},backref=false,colorlinks=false]
 {hyperref}
\hypersetup{pdftitle={Your Title},
 pdfauthor={Your Name},
 pdfpagelayout=OneColumn, pdfnewwindow=true, pdfstartview=XYZ, plainpages=false}

\makeatletter

\providecommand{\tabularnewline}{\\}

\IEEEoverridecommandlockouts
\ifCLASSOPTIONcompsoc
\usepackage[caption=false,font=normalsize,labelfont=sf,textfont=sf]{subfig}
\else
\usepackage[caption=false,font=footnotesize]{subfig}
\fi
\usepackage{cite}
\usepackage{amsthm}

\newtheorem{definition}{\textbf{Definition}}
\usepackage{algorithm}
\usepackage{algorithmic}
\usepackage{amsmath}

\@ifundefined{showcaptionsetup}{}{%
 \PassOptionsToPackage{caption=false}{subfig}}
\usepackage{subfig}
\makeatother

\begin{document}
\title{Efficient Betweenness Based Content Caching and Delivery Strategy
in Wireless Networks}
\author{\IEEEauthorblockN{Chenxi Zhao, Junyu Liu, Min Sheng, Yanpeng Dai}\\
\IEEEauthorblockA{State Key Laboratory of ISN, Xidian University,
Xi'an, Shaanxi, 710071, China}}
\maketitle
\begin{abstract}
In this work, we propose a content caching and delivery strategy to
maximize throughput capacity in cache-enabled wireless networks. To
this end, efficient betweenness (EB), which indicates the ratio of
content delivery paths passing through a node, is first defined to
capture the impact of content caching and delivery on network traffic
load distribution. Aided by EB, throughput capacity is shown to be
upper bounded by the minimal ratio of successful delivery probability
(SDP) to EB among all nodes. Through effectively matching nodes' EB
with their SDP, the proposed strategy improves throughput capacity
with low computation complexity. Simulation results show that the
gap between the proposed strategy and the optimal one (obtained through
exhausted search) is kept smaller than 6\%.
\end{abstract}

\section{Introduction}

In wireless networks, e.g., Internet of Things, sensor networks, etc.,
unbalanced network traffic load (NTL) distribution can result in congestion
on some nodes. The congestion will quickly spread to the entire network
and network throughput consequently decreases to be zero \cite{Wang2009Abrupt}.
Worse still, the unevenness of NTL distribution becomes more conspicuous
supposing that contents are pre-cached in nodes. Therefore, it is
critical to consider the impact of content caching and delivery on
NTL distribution to improve network throughput.

\begin{figure}[t]
\begin{centering}
\textsf{\includegraphics[scale=0.5]{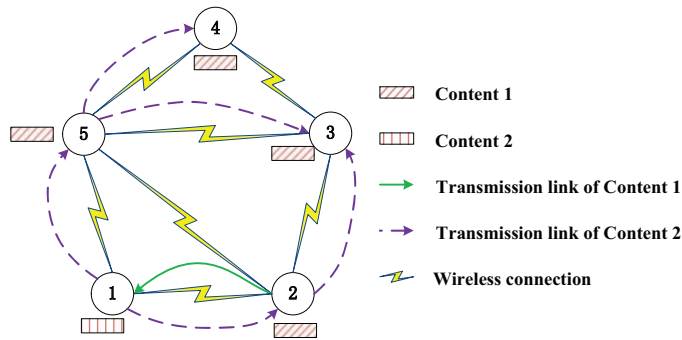}}
\par\end{centering}
\caption{\label{fig:FigOfEb}Illustration of content caching and delivery strategy.
There are five nodes and two contents in the network. Content caching
and delivery strategy is shown in the figure. Specifically, nodes
2, 3, 4, 5 cache content 1, and node 1 caches content 2. Moreover,
nodes 2, 3, 4, 5 request content 2 from node 1, and node 1 requests
content 1 from node 2.}
\end{figure}

Given no content is pre-cached, the traffic model in wireless networks
is generally modeled with random traffic. Particularly, per unit time,
each node generates contents in the same probability and sends each
content to a random destination. In this case, betweenness based approaches
are widely used to quantify NTL distribution \cite{betweennessDefination}.
Specifically, betweenness is defined as \cite{betweennessDefination}
\begin{equation}
b_{i}=\underset{j\neq k\neq i}{\sum}\frac{\phi\left(j,k,i\right)}{\phi\left(j,k\right)},\label{eq:DefOfBetweenness-1}
\end{equation}
where $\phi\left(j,k\right)$ denotes the number of the minimum-hop
paths from node $j$ to node $k$ and $\phi\left(j,k,i\right)$ denotes
the number of above paths that pass through node $i$. We give an
example in Fig. \ref{fig:FigOfEb}. Denote $b_{i}$ as the betweenness
of node $i$. Following the definition in \cite{betweennessDefination},
we have $b_{1}=0$, $b_{2}=1$, $b_{3}=1$, $b_{4}=0$ and $b_{5}=4$.
However, when contents are pre-cached in nodes, it is improper to
directly apply the random traffic model. Specifically, instead of
being delivered among all nodes, content is only delivered from nodes
caching content to nodes requesting content. In this case, betweenness
based approaches are obviously unsuitable for quantifying NTL distribution
in cache-enabled wireless networks (CWN). Therefore, how to quantify
NTL distribution in CWN remains to be investigated. 

\begin{figure}[t]
\begin{centering}
\textsf{\includegraphics[scale=0.5]{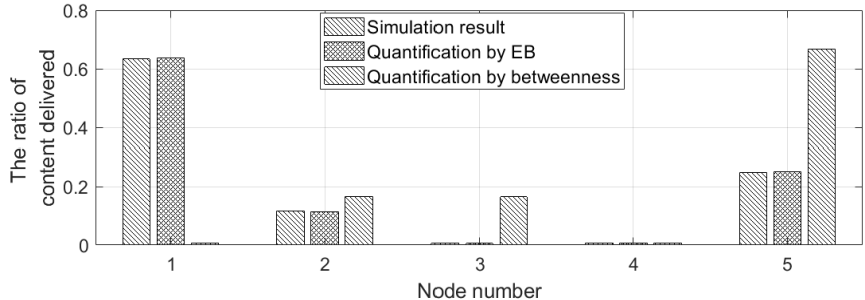}}
\par\end{centering}
\caption{\label{fig:FigOfCompare}The ratio of contents delivered by each node
under different quantification approaches. We set $q_{1}=q_{2}=0.5$.}
\end{figure}

To accurately characterize the NTL distribution in CWN, we propose
to define efficient betweenness (EB), which indicates the ratio of
content delivery paths passing through a node. Specifically, EB of
node $i$ is given by 
\begin{equation}
b_{i}^{\mathrm{E}}=\underset{s}{\sum}q_{s}\varphi\left(i,s\right).\label{eq:DefinationOfEB}
\end{equation}
In (\ref{eq:DefinationOfEB}), $\varphi\left(i,s\right)$ is the ratio
of the number of paths delivering content $s$ passing through node
$i$ to the total number of paths delivering content $s$. Moreover,
$q_{s}$ is denoted as the requested probability of content $s$.
According to (\ref{eq:DefinationOfEB}), \emph{it is shown that the
EB is proposed to capture the impact of content delivery processes
among nodes rather than all paths among nodes on NTL distribution.}
In the example in Fig. \ref{fig:FigOfEb}, we set the probability
of each node generating a content request as $\lambda_{1}=0.1$, $\lambda_{2}=0.2$,
$\lambda_{3}=0.3$, $\lambda_{4}=0.4$, and $\lambda_{5}=0.5$. Since
content 2 could be delivered from node 1 to node 3 by two content
delivery paths and one path passes through node 5, the average number
of paths delivering content 2 from node 1 to node 3 passing through
node 5 equals $0.5\lambda_{3}q_{2}$. Similarly, the number of paths
delivering content 2 from node 1 to node 4 passing through node 5
equals $\lambda_{4}q_{2}$. Therefore, we have $\varphi\left(5,2\right)=\frac{1}{\sum_{k=1}^{5}\lambda_{k}q_{2}}\left(0.5\lambda_{3}q_{2}+\lambda_{4}q_{2}\right)=\frac{11}{30}$.
Moreover, there is no path delivering content 1 passing through node
5. Hence, EB of node 5 is $b_{5}^{\mathrm{E}}=q_{2}\varphi\left(5,2\right)=\frac{11}{30}q_{2}$.
Similarly, we can obtain that $b_{1}^{\mathrm{E}}=\frac{14}{15}q_{2}$,
$b_{2}^{\mathrm{E}}=\frac{1}{15}q_{1}+0.1q_{2}$, $b_{3}^{\mathrm{E}}=0$,
and $b_{4}^{\mathrm{E}}=0$. To verify the accuracy of EB in quantifying
NTL distribution, we provide simulation results of NTL distribution
in Fig. \ref{fig:FigOfCompare}. Compared with the betweenness based
approach, it is obvious that the NTL distribution obtained through
the EB based approach is consistent with the real NTL distribution.

In this work, we propose an EB based content caching and delivery
strategy to improve throughput capacity in CWN. We formulate an optimization
problem to maximize the throughput capacity. Aided by the EB based
approach, the throughput capacity is shown to be upper bounded by
the minimum ratio of successful delivery probability (SDP) to EB among
all nodes. In particular, the throughput capacity is greatly degraded
supposing that low-SDP nodes have greater EB, which may consequently
result in network congestion. Therefore, we jointly design the content
caching and delivery to match an appropriate EB to the SDP of each
node. Due to the non-convexity of the primal problem, we solve it
by semidefinite relaxation and convex-concave procedure method. Numerical
results show that the gap between the proposed strategy and the optimal
one (obtained through exhausted search) is kept smaller than 6\% and
throughput capacity is improved by 60\% against the betweenness based
strategy under strong interference. Moreover, the computation complexity
is shown to be significantly reduced by the proposed algorithm, compared
to the optimal one.

\section{System Model}

\subsection{Network Model}

We consider a wireless network consisting of $N_{\mathrm{N}}$ nodes
in a two-dimensional plane. The set of nodes is denoted by $\mathcal{U}=\{1,...,N_{\mathrm{N}}\}$.
We use Voronoi Tessellation to divide the plane into small areas \cite{HetNets}.
Nodes located in adjacent areas can deliver contents directly. Let
$\mathcal{\mathit{\Psi}}=\{1,...,C\}$ denote the set of contents.
We assume that each content is of equal size, which is normalized
to one unit, and each node can at most cache $S$ contents. Denote
$q_{s}$ as the requested probability of content $s$, which follows
the Zipf distribution with parameter $\beta$ \cite{HetNets}. Particularly,
the requested probability of content $s$ is given by 
\begin{equation}
q_{s}=\frac{\frac{1}{s^{\beta}}}{\underset{t\in\mathcal{\mathit{\Psi}}}{\sum}\frac{1}{t^{\beta}}},
\end{equation}
where a large $\beta$ means that a few contents are requested by
majority of nodes. Moreover, we assume that all nodes share $N_{\mathrm{S}}$
subcarriers and each node randomly and independently occupies at most
one subcarrier. One node can successfully receive contents when received
signal-to-interference-noise-ratio (SINR) is greater than the predefined
SINR threshold \cite{interference}.

In this work, we define the throughput capacity by following the definition
in \cite{ThroughputDef}.

\begin{definition}[Throughput Capacity] For a given channel scheduling,
a throughput of $\lambda_{i}$ contents per unit time for node $i$
is feasible if node $i$ can receive $\lambda_{i}$ contents requested
by itself per unit time on average. In this context, the throughput
capacity of network is defined as $\sum_{i\in\mathcal{U}}\lambda_{i}$.\end{definition}

\subsection{Content Transmission Model \label{subsec:Content-transmission-Model}}

We consider that each node has a first-in-first-out transmission queue
with limited buffer size. Each content will be deleted from transmission
queues until it is successfully received by the next-hop node. Let
$R$ denote the transmission rate of each node, which is defined as
the maximal rate of transmitting contents by nodes. If the received
SINR is greater than the predefined SINR threshold, nodes can transmit
contents with the transmission rate. Moreover, contents are allowed
to be delivered among nodes via multi-hop paths and minimum-hop routing
is applied as the routing strategy in the CWN. If there are many minimum-hop
paths, the router will randomly choose one for each content.

\section{Throughput Capacity Maximization}

\subsection{NTL Distribution in CWN\label{subsec:Efficient-betweenness}}

In the following, we propose an EB based approach to quantify the
NTL distribution in CWN.

Let $x_{i,s}\subseteq\{0,1\}$ denote whether content $s$ is cached
in node $i$ or not, where $x_{i,s}=1$ indicates that node $i$ caches
content $s$ and $x_{i,s}=0$ otherwise. The number of contents cached
in each node should not be greater than $S$ and each content should
be cached once at least in the network\footnote{Note that we could first choose a subset of contents to cache in the
network if \textit{$C>N_{\mathrm{N}}*S$, }where the size of the subset
is not larger than $N_{\mathrm{N}}*S$. The difference of content
subset does not influence the analysis and proposed algorithm in the
following.}. Hence, we have
\begin{equation}
x_{i,s}\left(x_{i,s}-1\right)=0,\,\,\forall i\in\mathcal{U},\forall s\in\mathcal{\mathcal{\mathit{\Psi}}}\label{eq:Constraint3}
\end{equation}
\begin{equation}
\sum_{s\in\Psi}x_{i,s}\leq S,\,\,\forall i\in\mathcal{U}\label{eq:Constraint1}
\end{equation}
\begin{equation}
\sum_{i\in\mathcal{U}}x_{i,s}\geq1,\,\,\forall s\in\mathit{\Psi}.\label{eq:Constraint2}
\end{equation}
Meanwhile, we define a binary variable $y_{i,s,j}\subseteq\{0,1\}$,
where $y_{i,s,j}=1$ indicates that node $j$ requests content $s$
from node $i$ and $y_{i,s,j}=0$ otherwise. Moreover, we assume that
node $j$ requests content $s$ from only one of nodes caching content
$s$ and accordingly we have, 
\begin{equation}
y_{i,s,j}\left(y_{i,s,j}-1\right)=0,\,\,\forall i,j\in\mathcal{U},\forall s\in\mathit{\Psi}\label{eq:Constraint4}
\end{equation}
\begin{equation}
\sum_{i\in\mathcal{U}}y_{i,s,j}=1,\,\,\forall j\in\mathcal{U},\forall s\in\mathit{\Psi}.\label{eq:AccessOnlyOneNode}
\end{equation}
Therefore, $\varphi\left(i,s\right)$ in definition (\ref{eq:DefinationOfEB})
can be written as 
\begin{equation}
\varphi\left(i,s\right)=\frac{\sum_{j,k\in\mathcal{U}}\lambda_{k}q_{s}Ty_{j,s,k}\frac{\phi\left(j,k,i\right)}{\phi\left(j,k\right)}}{\sum_{k\in\mathcal{U}}\lambda_{k}q_{s}T}.\label{eq:Def of phi_i_s}
\end{equation}
In (\ref{eq:Def of phi_i_s}), since each content request will activate
a path to delivery the requested content, $\lambda_{k}q_{s}T$ is
equivalent to the average number of paths delivering content $s$
to node $k$ over $T$ seconds. Hence, $\sum_{k\in\mathcal{U}}\lambda_{k}q_{s}T$
is the total number of paths delivering content $s$ over $T$ seconds.
Moreover, $y_{j,s,k}\frac{\phi\left(j,k,i\right)}{\phi\left(j,k\right)}$
is the probability that paths delivering content $s$ from node $j$
to node $k$ pass through node $i$. Thus, $\lambda_{k}q_{s}T\sum_{j\in\mathcal{U}}y_{j,s,k}\frac{\phi\left(j,k,i\right)}{\phi\left(j,k\right)}$
is the average number of paths passing through node $i$ and delivering
content $s$ to node $k$ over $T$ seconds. Therefore, $\sum_{k\in\mathcal{U}}\lambda_{k}q_{s}T\sum_{j\in\mathcal{U}}y_{j,s,k}\frac{\phi\left(j,k,i\right)}{\phi\left(j,k\right)}$
is the average number of paths delivering content $s$ through node
$i$ over $T$ seconds. Moreover, we can define the average length
of all content delivery paths as 
\begin{equation}
L=\frac{1}{\sum_{i\in\mathcal{U}}\lambda_{i}T}\left(\sum_{k\in\mathcal{U}}\sum_{s\in\mathit{\Psi}}\lambda_{k}q_{s}T\sum_{j\in\mathcal{U}}y_{j,s,k}l_{j,k}\right),\label{eq:AveLength}
\end{equation}
where $l_{j,k}$ is the number of hops of the minimum-hop path between
node $j$ and node $k$. In (\ref{eq:AveLength}), $\sum_{i\in\mathcal{U}}\lambda_{i}T$
is the total number of paths delivering content over $T$ seconds.
Moreover, $\sum_{j\in\mathcal{U}}y_{j,s,k}l_{j,k}$ is the length
of the path delivering content $s$ to node $k$, and $\lambda_{k}q_{s}T$
is the number of times that node $k$ requests content $s$ over $T$
second. Hence, $\sum_{k\in\mathcal{U}}\sum_{s\in\mathit{\Psi}}\lambda_{k}q_{s}T\sum_{j\in\mathcal{U}}y_{j,s,k}l_{j,k}$
is the total length of all paths delivering content over $T$ seconds.
Furthermore, since node $j$ can request content $s$ from node $i$
only if node $i$ caches content $s$, $x_{i,s}$ and $y_{i,s,j}$
should satisfy 
\begin{equation}
y_{i,s,j}\leq x_{i,s},\,\,\forall i,j\in\mathcal{U},\forall s\in\mathit{\Psi}.\label{eq:ConstraintOfTwoVariables}
\end{equation}

\subsection{Throughput Capacity}

Denote $N_{all}$ as the total number of contents delivered to next-hop
nodes per unit time in the network. We first focus on the process
that node $j$ delivers content $s$ to node $k$. Denote $\left\{ n_{1},n_{2},...,n_{l_{j,k}},n_{l_{j,k}+1}\right\} $
as the node sequence of the minimum-hop path from node $j$ to node
$k$, where $n_{1}=j$ and $n_{l_{j,k}+1}=k$. If node $k$ requests
content $s$ from node $j$, the average number of content $s$ put
into the transmission queue of node $j$ per unit time equals $\lambda_{k}q_{s}$.
When the network goes into the steady state, the number of contents
put into the transmission queue of each node should equals the number
of delivered by this node. Hence, the average number of content $s$
delivered from node $n_{1}$ to node $n_{2}$ per unit time should
be equivalent to $\lambda_{k}q_{s}$. Similarly, for $\forall c\in\left\{ 1,...,l_{j,k}\right\} $,
the average number of content $s$ delivered from node $n_{c}$ to
node $n_{c+1}$ per unit time should be equivalent to $\lambda_{k}q_{s}$.
Therefore, the total number of content $s$ delivered by all nodes
in this path to their next-hop nodes per unit time equals $\lambda_{k}q_{s}l_{j,k}$.
Then, for all processes of content delivering, we can obtain $N_{all}$
by summing over $j$, $k$, and $s$
\begin{align*}
N_{all} & =\sum_{j\in\mathcal{U}}\sum_{k\in\mathcal{U}}\sum_{s\in\mathit{\Psi}}\lambda_{k}q_{s}y_{j,s,k}l_{j,k}\\
 & =L\sum_{i\in\mathcal{U}}\lambda_{i}.
\end{align*}
Moreover, since contents are delivered along the content delivery
paths from the source to the destination, the ratio of contents passing
through a node equals the ratio of content delivery paths passing
through a node, i.e., EB of this node. Hence, the probability that
contents will pass through a node equals the normalized EB of this
node, i.e., $b_{\mathrm{i}}^{\mathrm{E}}/\sum_{j\in\mathcal{U}}b_{j}^{\mathrm{E}}$
for node $i$, which is also verified in Fig. \ref{fig:FigOfCompare}.
Therefore, the number of contents that are put into the transmission
queue of node $i$ per unit time is given by
\begin{align}
N_{\mathrm{into}}\left(i\right) & =N_{all}\frac{b_{i}^{\mathrm{E}}}{\underset{j\in\mathcal{U}}{\sum}b_{j}^{\mathrm{E}}}\nonumber \\
 & =L\sum_{k\in\mathcal{U}}\lambda_{k}\frac{b_{i}^{\mathrm{E}}}{\underset{j\in\mathcal{U}}{\sum}b_{j}^{\mathrm{E}}}.\label{eq:NumberOfInto}
\end{align}
We assume that there are $M$ minimum-hop paths from node $j$ to
node $k$. Denote $\left\{ m_{1},m_{2},...,m_{l_{j,k}},m_{l_{j,k}+1}\right\} $
as the node sequence of the $m$-th minimum-hop path, where $m_{i}$
denotes the $i$-th node in the $m$-th minimum-hop path. Then, we
have 
\begin{align}
\sum_{i\in\mathcal{U\backslash\left\{ \mathit{k}\right\} }}\frac{\phi\left(j,k,i\right)}{\phi\left(j,k\right)} & =\sum_{h=1}^{l_{j,k}}\sum_{m=1}^{M}\frac{\phi\left(j,k,m_{h}\right)}{\phi\left(j,k\right)}\nonumber \\
 & =\sum_{h=1}^{l_{j,k}}\frac{\phi\left(j,k\right)}{\phi\left(j,k\right)}\nonumber \\
 & =l_{j,k}.
\end{align}
Therefore, the sum of all nodes' EB can be written as
\begin{align}
\underset{i\in\mathcal{U}}{\sum}b_{i}^{\mathrm{E}} & =\underset{i\in\mathcal{U}\backslash\left\{ \mathit{k}\right\} }{\sum}\underset{s\in\mathcal{\mathit{\Psi}}}{\sum}q_{s}\frac{\sum_{j,k\in\mathcal{U}}\lambda_{k}q_{s}Ty_{j,s,k}\frac{\phi\left(j,k,i\right)}{\phi\left(j,k\right)}}{\sum_{k\in\mathcal{U}}\lambda_{k}q_{s}T}\nonumber \\
 & =\underset{s\in\mathcal{\mathit{\Psi}}}{\sum}q_{s}\frac{\sum_{j,k\in\mathcal{U}}\lambda_{k}y_{j,s,k}\underset{i\in\mathcal{U}\backslash\left\{ \mathit{k}\right\} }{\sum}\frac{\phi\left(j,k,i\right)}{\phi\left(j,k\right)}}{\sum_{k\in\mathcal{U}}\lambda_{k}}\nonumber \\
 & =\underset{s\in\mathcal{\mathit{\Psi}}}{\sum}q_{s}\frac{\sum_{j,k\in\mathcal{U}}\lambda_{k}y_{j,s,k}l_{j,k}}{\sum_{k\in\mathcal{U}}\lambda_{k}}\nonumber \\
 & =L.\label{eq:SumOfEB}
\end{align}

Substituting (\ref{eq:SumOfEB}) into (\ref{eq:NumberOfInto}), we
can obtain $N_{\mathrm{into}}\left(i\right)=\sum_{k\in\mathcal{U}}\lambda_{k}b_{i}^{\mathrm{E}}$.
Moreover, the number of contents that node $i$ can deliver successfully
per unit time is $N_{\mathrm{del}}\left(i\right)=p_{i}R$, where $p_{i}$
is the successful delivery probability (SDP) of node $i$. $N_{\mathrm{into}}\left(i\right)$
and $N_{\mathrm{del}}\left(i\right)$ should satisfy $N_{\mathrm{into}}\left(i\right)\leq N_{\mathrm{del}}\left(i\right)$
if there is no local congestion in node $i$. Hence, if no congestion
occurs in any node in the network, $\lambda_{i},\forall i\in\mathcal{U}$
should satisfy
\begin{equation}
\sum_{k\in\mathcal{U}}\lambda_{k}b_{i}^{\mathrm{E}}\leq p_{i}R,\,\,\forall i\in\mathcal{U}.\label{eq:NewConclusion}
\end{equation}
In this work, we consider the case that each node has the same probability
of generating a content request per unit time, namely, $\lambda_{i}=\lambda,\,\forall i\in\mathcal{U}$.
In this case, (\ref{eq:NewConclusion}) can be rewritten as 
\begin{equation}
\lambda\leq\underset{i\in\mathcal{U}}{\min}\,\frac{p_{i}R}{N_{\mathrm{N}}b_{i}^{\mathrm{E}}}.\label{eq:DefinationOfLamda}
\end{equation}
According to the Definition 1, the throughput capacity of network
equals $\lambda N_{\mathrm{N}}$. According to (\ref{eq:DefinationOfLamda}),
we can obtain that the upper bound of throughput capacity is $\Theta\left(\min_{i}\left(p_{i}/b_{i}^{\mathrm{E}}\right)\right)$
under given network parameters. Therefore, we could match the NTL
distribution with SDP to maximize throughput capacity.

\subsection{EB Based Strategy}

In the following, we elaborate the detail of the EB based content
caching and delivery strategy (ECCDS).

Firstly, we formulate the following optimization problem
\begin{align*}
(\mathrm{P0}):\underset{}{\mathrm{\max}} & \,\,\underset{i\in\mathcal{U}}{\min}\,\frac{p_{i}}{b_{i}^{\mathrm{E}}}\\
\mathrm{s.t.} & (\ref{eq:Constraint3})-(\ref{eq:ConstraintOfTwoVariables})
\end{align*}
To solve the problem (P0), we introduce a new variate $w$, which
satisfies constraint (\ref{eq:Constraint5}) 
\begin{equation}
w\geq\frac{b_{i}^{\mathrm{E}}}{p_{i}},\,\,\forall i\in\mathcal{U}.\label{eq:Constraint5}
\end{equation}
 Hence, problem (P0) can be transformed into (P1)
\begin{align*}
(\mathrm{P1}):\min & \,w\\
\mathrm{s.t.} & (\ref{eq:Constraint3})-(\ref{eq:AccessOnlyOneNode}),\left(\ref{eq:ConstraintOfTwoVariables}\right),(\ref{eq:Constraint5})
\end{align*}
The problem (P1) is an integer linear programming problem. Semidefinite
relaxation (SDR) approach \cite{Luo2010Semidefinite} can be applied
to solve (P1). Aided by SDR, the following relaxed problem can be
obtained, 
\begin{align}
(\mathrm{P2}):\min & \,w\nonumber \\
\mathrm{s.t.} & \,(\ref{eq:Constraint1}),(\ref{eq:Constraint2}),(\ref{eq:AccessOnlyOneNode}),(\ref{eq:ConstraintOfTwoVariables}),(\ref{eq:Constraint5})\nonumber \\
 & \,Tr(Q_{l}Z)-q_{l}\boldsymbol{z}=0\label{eq:SDR}
\end{align}
where $Z=\boldsymbol{\boldsymbol{z}}\boldsymbol{\boldsymbol{z}}{}^{T}$
and $\boldsymbol{\boldsymbol{z}}=\left[\boldsymbol{y},\boldsymbol{x}\right]^{\mathrm{T}}$,
with $\boldsymbol{y}=[y_{1,1,1},y_{1,2,1},...,y_{1,C,N_{\mathrm{N}}},...,y_{N_{\mathrm{N}},C,N_{\mathrm{N}}}]$
and $\boldsymbol{x}=[x_{1,1},x_{1,2},...,x_{1,C},...,x_{N_{\mathrm{N}},C}]$.
Moreover, in $\left(\ref{eq:SDR}\right)$, $q_{l}$ is a standard
unit vector with the $l$-th entry being 1 and $Q_{l}=\mathrm{diag}(q_{l})$
with $l\in\{1,...,(N_{\mathrm{N}}+1)N_{\mathrm{N}}C\}$. (P2) is a
convex programming which can be solved by using the convex optimization
toolbox, such as CVX \cite{cvx}. After obtaining an optimal solution
$\boldsymbol{\mathcal{Z}}^{*}$ to the problem (P2), we can generate
a series of random vectors $\xi_{l}\sim\mathcal{N}(0,\boldsymbol{\mathcal{Z}}^{*})$
as recovery samples. Then, we consider the penalty convex-concave
procedure (CCP) method to map each of these recovery samples to the
feasible set of problem (P1)\cite{Lipp2016Variations}. In each iteration,
we first define an initial vector $\boldsymbol{v}_{0}$, which is
one of the samples above. Then, the problem (P2) can be transformed
into (P3)
\begin{align}
(\mathrm{P3}):\underset{x}{\mathrm{\min}} & \,w+\tau_{a}\sum_{i=1}^{(N_{\mathrm{N}}+1)N_{\mathrm{N}}C}\omega_{i}\nonumber \\
s.t. & \,(\ref{eq:Constraint1}),(\ref{eq:Constraint2}),(\ref{eq:AccessOnlyOneNode}),(\ref{eq:ConstraintOfTwoVariables}),(\ref{eq:Constraint5})\nonumber \\
 & \,q_{l}\boldsymbol{\boldsymbol{z}}-\stackrel{\land}{g}_{k}(\boldsymbol{\boldsymbol{z}};\boldsymbol{\boldsymbol{z}}_{a})\leq\boldsymbol{w}\\
 & \,z_{i}\in[0,1],\,\forall i\in\{1,...,(N_{\mathrm{N}}+1)N_{\mathrm{N}}C\}\\
 & \,w_{i}>0,\,\forall i\in\{1,...,(N_{\mathrm{N}}+1)N_{\mathrm{N}}C\}
\end{align}
where $\stackrel{\land}{g}_{k}(\boldsymbol{\boldsymbol{z}};\boldsymbol{\boldsymbol{z}}_{a})=g_{k}(\boldsymbol{\boldsymbol{z}}_{a})+\nabla g_{k}(\boldsymbol{\boldsymbol{z}}_{a})^{T}(\boldsymbol{\boldsymbol{z}}-\boldsymbol{\boldsymbol{z}}_{a})$
with $g_{k}(\boldsymbol{\boldsymbol{z}}_{a})=\boldsymbol{\boldsymbol{z}}_{a}^{T}Q_{l}\boldsymbol{\boldsymbol{z}}_{a}$,
$\boldsymbol{w}=(w_{1},...,w_{(N_{\mathrm{N}}+1)N_{\mathrm{N}}C})$,
and $\boldsymbol{\boldsymbol{z}}=\left(z_{1},...,z_{(N_{\mathrm{N}}+1)N_{\mathrm{N}}C}\right)$.
We substitute $\boldsymbol{\boldsymbol{z}}_{a}=\boldsymbol{v}_{0}$
into (P3). After obtaining the solution $\boldsymbol{v}^{'}$, let
$\boldsymbol{v}_{a}=\boldsymbol{v}^{'}$ and increase the penalty
factor $\tau_{a}$. Substitute iteratively $\boldsymbol{v}_{a}$ and
$\tau_{a}$ into (P3) until stopping criterion is satisfied. The algorithm
of penalty CCP method is written in Algorithm 1.

\subsubsection*{\textsl{Computation complexity analysis}}

The proposed algorithm contains two parts, namely, the SDR algorithm
and the penalty CCP algorithm. The computation complexity of the two
algorithms are studied in the following. (1) The SDR relaxed problem
of problem (P1) can be solved by the interior-point algorithm with
a worst case complexity of $O\left(\max\left\{ \theta_{p},\theta_{c}\right\} ^{4}\theta_{p}^{0.5}\log\left(1/\epsilon\right)\right)$
given a solution accuracy $\epsilon>0$ \cite{Complexity1}. Specifically,
$\theta_{p}$ is the number of variables in problem (P1) and $\theta_{c}$
is the number of constrains in problem (P1). Furthermore, the SDR
complexity scales slowly (logarithmically) with $\epsilon$. (2) problem
(P3) can be solved by the interior-point algorithm with a worst case
complexity of $O\left(\left(\theta_{p}+\theta_{c}\right)^{3.5}\right)$
\cite{Complexity2}. Therefore, the computational complexity of the
penalty CCP algorithm is $O\left(l\theta_{r}\left(\theta_{p}+\theta_{c}\right)^{3.5}\right)$.
Specifically, $l$ is the number of recovery samples and $\theta_{r}$
is the number of iteration for each recovery sample. Therefore, the
computational complexity of the proposed algorithm is $O\left(\max\left\{ \theta_{p},\theta_{c}\right\} ^{4}\theta_{p}^{0.5}\log\left(1/\epsilon\right)+l\theta_{r}\left(\theta_{p}+\theta_{c}\right)^{3.5}\right)$.
For comparison, we also give the computational complexity of the optimal
approach, e.g., branch and bound (B\&B) method. The worst case of
B\&B is that each device tries all the possible set of content to
cache and tries all the possible set of device to request. The computational
complexity of content caching process is $O\left(\left(\frac{C!}{\left(C-S\right)!}\right)^{N_{n}}\right)$.
Similarly, the computational complexity of content caching process
is $O\left(\left(N_{n}C\right)^{N_{n}}\right)$. Therefore, the computational
complexity of B\&B method is $O\left(\left(\frac{C!}{\left(C-S\right)!}\right)^{N_{n}}\left(N_{n}C\right)^{N_{n}}\right)$.
Comparing with the optimal approach, the proposed algorithm can simplify
the computation complexity from a power function of the problem size
and the number of constraints to a polynomial function. Therefore,
the proposed algorithm is a computationally efficient approximation
approach to solve problem (P1).

\section{Simulation Results}

\floatname{algorithm}{Algorithm}
\begin{algorithm}[!t]
\caption{The penalty CCP algorithm.}  
\begin{algorithmic}[1] \label{algorithm:recovery algorithm}

\STATE \textbf{Initialization}
\STATE $\quad\bullet$ Given an SDR solution $\mathcal{X}^{*}$.
\STATE $\quad\bullet$ Given the number of random samples $\mathcal{L}$ and set $l=1$.

\REPEAT
\STATE Generation $\zeta_{l}\sim\mathcal{N}(0,\mathcal{X}^{*})$ and set $a=0$.
\STATE Given initial point $v_{0}=\zeta_{l}$, penalty paramaters $\tau_{0}>0$, $\tau_{\max}>0$, and $\theta>1$.
\REPEAT
\STATE According $v_{a}$ and $\tau_{a}$ form (P3).
\STATE Set the value of $v_{a+1}$ to a solution of (P3).
\STATE Increase penalty factor $\tau_{a+1}=\mathrm{min( \theta\tau_{a},\tau_{max})}$.
\STATE $a=a+1$.
\UNTIL{$\tau_{a}=\tau_{max}$}

\STATE $l=l+1$.
\UNTIL{$l=\mathcal{L}+1$}
\STATE $l^{*}=\mathrm{arg}\,\mathrm{max\left\{ \mathit{q}\mathit{v}_{\mathit{l}}\right\} }, \forall\mathit{l}\in\left\{1,...,\mathcal{L}\right\}$.

\RETURN $\boldsymbol{\boldsymbol{z}}^{*}=\boldsymbol{\boldsymbol{z}}_{l^{*}}$
\end{algorithmic} 
\end{algorithm}

In this section, the simulation results are given. We consider that
the number of nodes and contents both equal 10 ($N_{\mathrm{N}}=10$,
$C=10$). All nodes are independently located within an area of the
100$\times$100 square meters according to uniform distribution. Moreover,
it is assumed that each content is of identical size and, for simplicity,
the content size is set to be 1. Note that the results and proposed
strategy could be readily applied and extended supposing that the
contents are of different size, since they could partitioned into
chunks of the identical size \cite{ContentSize}. We consider the
transmission power of nodes as $P=20\mathrm{dBm}$. The pathloss exponent
and noise coefficient are set as $\alpha=4$ and $\sigma^{2}=-120\mathrm{dBm}$,
respectively.

\begin{figure}[t]
\subfloat[\label{fig:Sim v.s. Ana}]{\centering{}\includegraphics[scale=0.5]{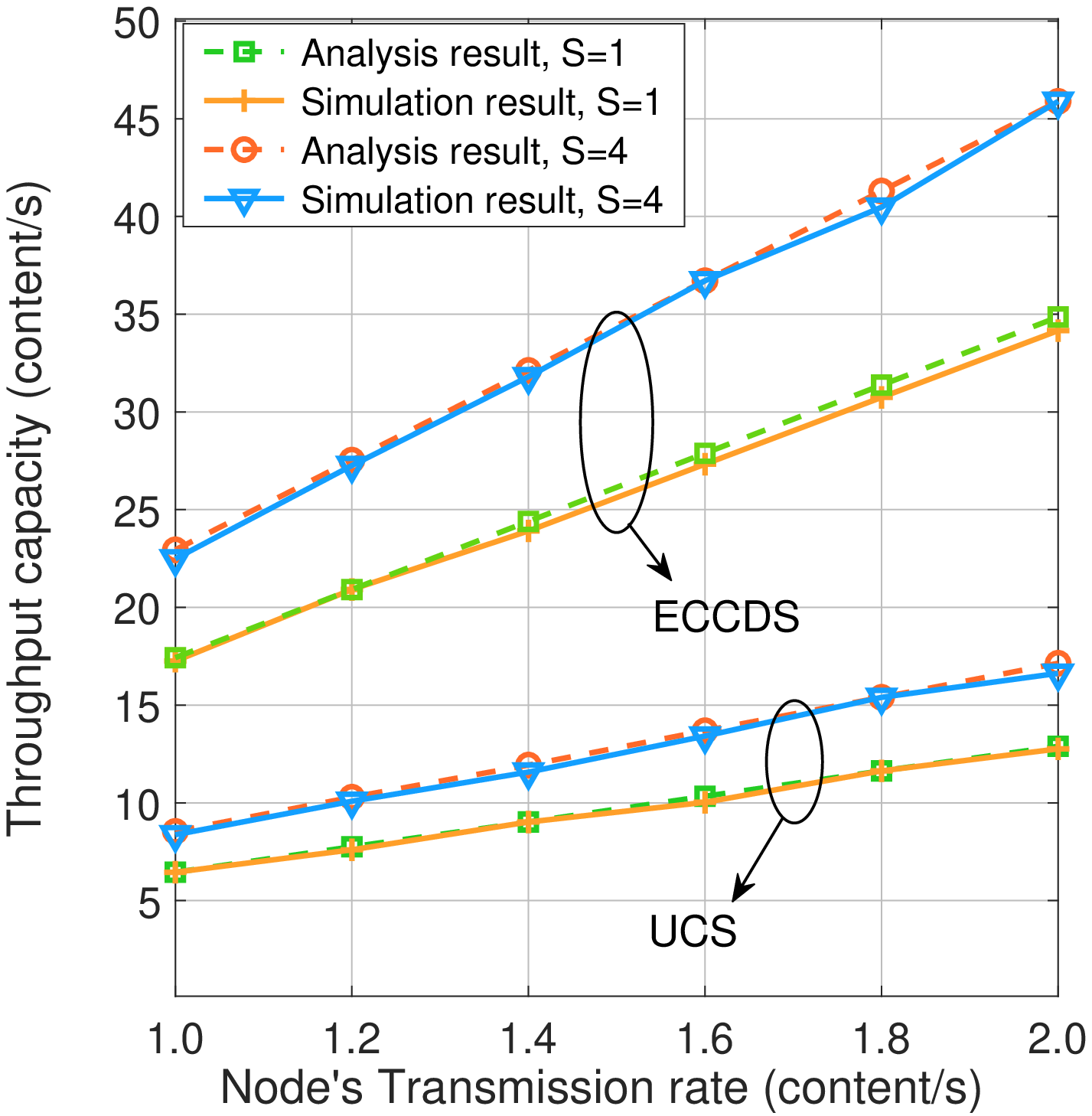}}\subfloat[\label{fig:NTCCompare}]{\begin{centering}
\textsf{\includegraphics[scale=0.5]{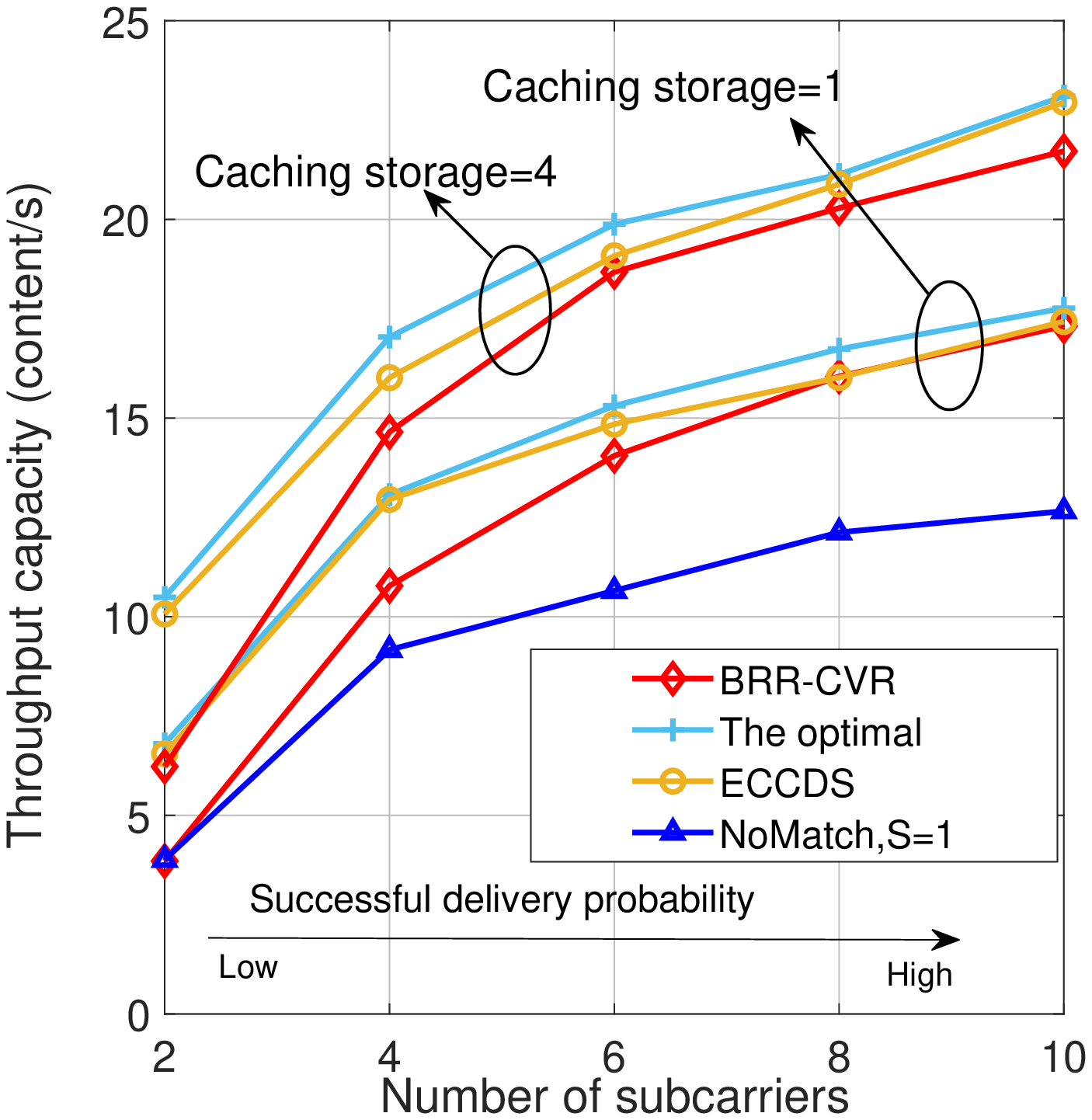}}
\par\end{centering}
}

\caption{(a) Throughput capacity versus node's transmission rate with setting
$\tau=3\mathrm{dB}$, $N_{\mathrm{S}}=10$, and $\beta=1$. (b) Throughput
capacity versus the number of subcarriers. We set the system setting
as $\tau=3\mathrm{dB}$, $\beta=1$, and $R=1\,(content/s)$.}
\end{figure}

Fig. \ref{fig:Sim v.s. Ana} shows the performance gap between analysis
and simulation for two strategies, i.e., ECCDS and uniform caching
strategy (UCS). Specifically, in UCS, each node caches each content
with same probability. We can observe that simulation results match
well with analysis results, thus validating our analysis. Moreover,
in Table \ref{tab:Throughput v.s. N_s}, we give simulation results
of throughput capacity under different the number of subcarriers $\left(N_{\mathrm{S}}\right)$.
We set the node's transmission rate $R=2\,(content/s)$, the SINR
threshold $\tau=3\mathrm{dB}$, and caching storage $S=4$. We can
observe that the value of $\left(\lambda/\min_{i}\left(p_{i}/b_{i}^{\mathrm{E}}\right)\right)$
equals a constant under different $N_{\mathrm{S}}$. In other words,
the throughput capacity is $\Theta\left(\min_{i}\left(p_{i}/b_{i}^{\mathrm{E}}\right)\right)$.

\begin{table}[t]
\caption{\label{tab:Throughput v.s. N_s}Throughput capacity versus $\min_{i}\left(p_{i}/b_{i}^{\mathrm{E}}\right)$}

\centering{}%
\begin{tabular}{|c|c|c|c|}
\hline 
$N_{\mathrm{S}}$ & Throughput capacity $\left(\lambda N_{\mathrm{N}}\right)$ & $\min_{i}\left(p_{i}/b_{i}^{\mathrm{E}}\right)$ & $\frac{\lambda N_{\mathrm{N}}}{\min_{i}\left(p_{i}/b_{i}^{\mathrm{E}}\right)}$\tabularnewline
\hline 
\hline 
2 & 20.1322 & 10.0651 & 2.0002\tabularnewline
\hline 
4 & 32.0327 & 16.0172 & 1.9999\tabularnewline
\hline 
6 & 38.1624 & 19.0850 & 1.9996\tabularnewline
\hline 
8 & 41.7641 & 20.8790 & 2.0003\tabularnewline
\hline 
10 & 45.8914 & 22.9446 & 2.0001\tabularnewline
\hline 
\end{tabular}
\end{table}

We then compare ECCDS with the BRR-CVR strategy \cite{BetweennessCachingWireless}.
BRR-CVR strategy is a betweenness based caching strategy, where the
popular contents are cached in nodes with high betweenness. We plot
throughput capacity as a function of the number of subcarriers in
Fig. \ref{fig:NTCCompare}. For the BRR-CVR strategy, since betweenness
can not exactly quantify the NTL distribution in CWN, low-SDP nodes
may have high EB, which results in congestion and further degrades
the throughput capacity. Moreover, we also provide the throughput
capacity of the strategy without matching EB with SDP. The results
validates the necessity of matching nodes' EB with their SDP.

\section{Conclusion}

In this work, we investigate the content caching and delivery strategy
to improve throughput capacity of CWN. We define an efficient metric,
i.e., EB, to quantify the NTL distribution in CWN. Aided by the EB
based approach, we derive that the throughput capacity is upper bounded
by the minimum ratio of SDP to EB among all nodes. To maximize the
throughput capacity, we design the ECCDS to match an appropriate EB
for the SDP of each node. The simulation results show that, particularly
under strong interference, the ECCDS can efficiently improve the throughput
capacity.

\bibliographystyle{IEEEtran}
\bibliography{WCL}

\begin{thebibliography}{10}
\providecommand{\url}[1]{#1}
\csname url@samestyle\endcsname
\providecommand{\newblock}{\relax}
\providecommand{\bibinfo}[2]{#2}
\providecommand{\BIBentrySTDinterwordspacing}{\spaceskip=0pt\relax}
\providecommand{\BIBentryALTinterwordstretchfactor}{4}
\providecommand{\BIBentryALTinterwordspacing}{\spaceskip=\fontdimen2\font plus
\BIBentryALTinterwordstretchfactor\fontdimen3\font minus
  \fontdimen4\font\relax}
\providecommand{\BIBforeignlanguage}[2]{{%
\expandafter\ifx\csname l@#1\endcsname\relax
\typeout{** WARNING: IEEEtran.bst: No hyphenation pattern has been}%
\typeout{** loaded for the language `#1'. Using the pattern for}%
\typeout{** the default language instead.}%
\else
\language=\csname l@#1\endcsname
\fi
#2}}
\providecommand{\BIBdecl}{\relax}
\BIBdecl

\bibitem{Wang2009Abrupt}
W.~X. Wang, Z.~X. Wu, R.~Jiang, G.~Chen, and Y.~C. Lai, ``Abrupt transition to
  complete congestion on complex networks and control,'' \emph{Chaos}, vol.~19,
  no.~3, p. 824, 2009.

\bibitem{betweennessDefination}
S.~Boccaletti, V.~Latora, Y.~Moreno, M.~Chavez, and D.~.~U. Hwang, ``Complex
  networks: Structure and dynamics,'' \emph{PHYS REP}, vol. 424, no. 4-5, pp.
  175--308, Feb. 2006.

\bibitem{HetNets}
M.~{Mirahsan}, R.~{Schoenen}, and H.~{Yanikomeroglu}, ``{HetHetNets}:
  Heterogeneous traffic distribution in heterogeneous wireless cellular
  networks,'' \emph{IEEE J. Select. Areas Commun.}, vol.~33, no.~10, pp.
  2252--2265, Oct. 2015.

\bibitem{interference}
J.~{Liu}, M.~{Sheng}, L.~{Liu}, and J.~{Li}, ``Performance of small cell
  networks under multislope bounded pathloss model: From sparse to ultradense
  deployment,'' \emph{IEEE Trans. Veh. Technol.}, vol.~67, no.~11, pp.
  11\,022--11\,034, Nov. 2018.

\bibitem{ThroughputDef}
P.~Gupta and P.~R. Kumar, ``The capacity of wireless networks,'' \emph{IEEE
  Trans. Inform. Theory}, vol.~46, no.~2, pp. 388--404, 2000.

\bibitem{Luo2010Semidefinite}
Z.~Q. Luo, W.~K. Ma, M.~C. So, Y.~Ye, and S.~Zhang, ``Semidefinite relaxation
  of quadratic optimization problems,'' \emph{IEEE Signal Process. Mag.},
  vol.~27, no.~3, pp. 20--34, 2010.

\bibitem{cvx}
M.~Grant and S.~Boyd, ``{CVX}: Matlab software for disciplined convex
  programming, version 2.1,'' \url{http://cvxr.com/cvx}, Mar. 2014.

\bibitem{Lipp2016Variations}
T.~Lipp and S.~Boyd, ``Variations and extension of the convex-concave
  procedure,'' \emph{Optim. Eng.}, vol.~17, no.~2, pp. 263--287, 2016.

\bibitem{Complexity1}
Z.~{Luo}, W.~{Ma}, A.~M. {So}, Y.~{Ye}, and S.~{Zhang}, ``Semidefinite
  relaxation of quadratic optimization problems,'' \emph{IEEE Signal Process.
  Mag.}, vol.~27, no.~3, pp. 20--34, 2010.

\bibitem{Complexity2}
C.~{Qian}, N.~D. {Sidiropoulos}, K.~{Huang}, L.~{Huang}, and H.~C. {So},
  ``Phase retrieval using feasible point pursuit: Algorithms and cramer–rao
  bound,'' \emph{IEEE Trans. Signal Process.}, vol.~64, no.~20, pp. 5282--5296,
  2016.

\bibitem{ContentSize}
Z.~{Qin}, X.~{Gan}, L.~{Fu}, X.~{Di}, J.~{Tian}, and X.~{Wang}, ``Content
  delivery in cache-enabled wireless evolving social networks,'' \emph{IEEE
  Trans. Wireless Commun.}, vol.~17, no.~10, pp. 6749--6761, 2018.

\bibitem{BetweennessCachingWireless}
B.~{Chen}, L.~{Liu}, Z.~{Zhang}, W.~{Yang}, and H.~{Ma}, ``Brr-cvr: A
  collaborative caching strategy for information-centric wireless sensor
  networks,'' in \emph{Proc. IEEE 12th Int. Conf. Mobile Ad-Hoc Sensor Netw.
  (MSN)}, 2016, pp. 31--38.

\end{thebibliography}

\end{document}